
\input phyzzx

\sequentialequations

\overfullrule=0pt
\catcode`\@=11

\def \O{{\cal O}}
\def \CO{{\cal O}}

\def\NP{{\it Nucl. Phys.\ }}

\def\PL{{\it Phys. Lett.\ }}

\def\CMP{{\it Comm. Math. Phys.\ }}

\def\Mod{{\it Mod. Phys. Lett.\ }}

\def\CO{{\cal O}}

\def\eqaligntwo#1{\null\,\vcenter{\openup\jot\m@th
\ialign{\strut\hfil
$\displaystyle{##}$&$\displaystyle{{}##}$&$\displaystyle{{}##}$\hfil
\crcr#1\crcr}}\,}
\catcode`\@=12
\REF\GKN{D. J. Gross, I. R. Klebanov and M. J. Newman
\journal Nucl. Phys. &B350 (1991) 621.}
\REF\ferm{D. J. Gross and I. R. Klebanov
\journal Nucl. Phys. &B352 (1991) 671.}
\REF\Jev{K. Demeterfi, A. Jevicki and J. P. Rodrigues,
Brown preprints  BROWN-HET-795 and BROWN-HET-803 (1991).}
\REF\GD{U. H. Danielsson and D. J. Gross, Princeton preprint PUPT-1258.}
\REF\AMP{A. M. Polyakov, \Mod {\bf A6} (1991) 635.}
\REF\lz{B. Lian and G. Zuckerman, \PL {\bf 254B} (1991) 417;
Yale preprint YCTP-P18-91.}
\REF\BKl{M. Bershadsky and I. R. Klebanov \journal Nucl. Phys.
&B360 (1991) 559.}
\REF\imb{C. Imbimbo, S. Mahapatra and S. Mukhi,
Genova and Tata Inst. preprint GEF-TH 8/91, TIFR/TH/91-27.}
\REF\Ed{E. Witten, Inst. for Advanced Study preprint IASSNS-HEP-91/51.}
\REF\GKleb{D. J. Gross and I. R. Klebanov, \NP {\bf B359} (1991) 3.}
\REF\States{J. Goldstone, unpublished; V. G. Kac, in
{\sl Group Theoretical Methods in Physics}, Lecture Notes in Physics,
vol. 94 (Springer-Verlag, 1979).}
\REF\sasha{A. M. Polyakov, ZhETF 63 (1972) 24.}
\REF\bak{I. Bakas, \PL {\bf B228} (1989) 57.}
\REF\PRS{C. Pope, L. Romans and X. Shen, \NP {\bf B339} (1990) 191.}
\REF\Berg{E. Bergshoeff, M. P. Blencowe and K. S. Stelle,
\CMP {\bf 128} (1990) 213.}
\REF\avan{J. Avan and A. Jevicki, Brown preprints BROWN-HET-801
and BROWN-HET-824; M. Awada and S. J. Sin, Florida preprint UFITP-HEP-91-03;
A. Gerasimov, A. Marshakov, A. Mironov, A. Morozov, and A. Orlov,
Lebedev Inst. preprints;
D. Minic, J. Polchinski and Z. Yang, Univ. of Texas
preprint UTTG-16-91;
G. Moore and N. Seiberg, Rutgers and Yale preprint RU-91-29,
YCTP-P19-91.}

\def\CO{{\cal O}}
\def\eqaligntwo#1{\null\,\vcenter{\openup\jot\m@th
\ialign{\strut\hfil
$\displaystyle{##}$&$\displaystyle{{}##}$&$\displaystyle{{}##}$\hfil
\crcr#1\crcr}}\,}
\catcode`\@=12

\def\O{\Omega}

\def\qg{quantum gravity}

\Pubnum={PUPT-1281}
\date={September 1991}
\titlepage
\title{Interaction of Discrete States in Two-Dimensional String Theory.}
\author{I. R. Klebanov\foot{ Supported in part by DOE grant
DE-AC02-76WRO3072, A. P. Sloan Foundation,
and an NSF Presidential Young Investigator Award.}
 and A. M. Polyakov\foot{ Supported in part by NSF grant PHY90-21984.}
}
\JHL
\abstract
We study the couplings of discrete states that appear in the
string theory embedded in two dimensions, and show that they
are given by the structure constants of the group of area preserving
diffeomorphisms. We propose an effective action for these states,
which is itself invariant under this infinite-dimensional group.
\endpage

Recent investigations of two-dimensional \qg\ coupled to
simple matter systems have displayed very rich and fascinating
structures. It is clearly important to concentrate on those
of them that have potential significance beyond the particular
models with $c\leq 1$. The choice to be made is, of course,
a matter of taste. Our preference in this paper will be given
to the infinite sequence of discrete states that appeared
in the $c=1$ matrix model [\GKN-\GD],
and were found and interpreted
as higher string states in the continuum theory [\AMP].
\foot{Analyses of states involving ghost excitations have appeared
in ref. [\lz-\Ed]. In this letter we consider only the discrete
states with no ghost excitations.}
We will show here that these states interact with each other,
and that their couplings are given by the structure constants
of the area preserving diffeomorphisms. Moreover, the interaction
itself is invariant under this infinite-dimensional group.
Note that this is the symmetry group of the ideal incompressible
fluid, which is perhaps related to the string field theory in
our case.

To set the stage, let us remind how the discrete states appear in the
theory. The $c=1$ model is described by the two-dimensional
critical string with coordinates
$X^\mu= (\phi, X)$ and a linear dilaton background.
$$ S={1 \over 8\pi}\int d^2\sigma \biggl({2\over\alpha'}
(\partial_a X\partial^a X+\partial_a \phi\partial^a\phi)
-2b_\mu X^\mu R^{(2)}\biggr)-{1\over 2\pi}\oint ds \kappa b_\mu X^\mu
\eqn\eq$$
where $b_\mu=(2/\sqrt{\alpha'}, 0)$,
and the last term involving the boundary curvature $\kappa$ is
needed for the proper description of open strings.
Because of the background charge for the Liouville field,
the $SL(2, C)$ invariance is maintained in the shifted Koba-Nielsen
amplitudes, which are proportional to the volume of the Liouville zero
mode [\AMP, \GKleb].
In the discussion of closed strings we will
set $\alpha'=4$ while for open strings we will choose $\alpha'=1$.
The physical states are those that
satisfy the Virasoro conditions modulo gauge transformations.
The simplest non-trivial example is provided by the level one
(vector) operator of the open string
$\int ds e_\mu (p, \epsilon)\partial X^\mu e^{ipX}e^{\epsilon\phi}$.
The conditions that this operator is physical can be seen to be
$$\eqalign{&(f_\mu+b_\mu) e^\mu (f)=0\cr &
e_\mu (f)\sim e_\mu+\lambda f_\mu\cr
& f_\mu (f^\mu +b^\mu)= 0\cr }\eqn\vector
$$
Here $f=(\epsilon, p)$ is the two-dimensional momentum, with
$\epsilon$ being the Liouville energy.
The scalar products look like the usual Lorentzian ones,
$ f_\mu e^\mu=f^1 e^1 -f^0 e^0$.

For generic $p$ there are no solutions of eq. \vector. However,
it is obvious that for $f_\mu=0$ and for $f_\mu=-b_\mu$ there is a jump in
the number of degrees of freedom.
We thus find solutions with $p=0$ and
$\epsilon= 0$ or $\epsilon=-2$.
Simple analysis shows that
at these values one can chose the ``material gauge''
$b_\mu e^\mu=0$ or $e_0=0$.
Thus, these
vertex operators do not depend on the oscillators of $\phi$.
In fact, this property persists at all the mass levels. It can be
shown that for $c=1$ there exists a non-singular gauge such
that the oscillators of the Liouville field are not excited.
Therefore, all one should do in order to find the discrete
states is look for the Virasoro primary fields in the Fock space
of the $X$-field. Fortunately, the problem of finding all such
states has been solved long ago [\States]. Here we will summarize the results.

At a generic momentum $p$ the only Virasoro primary field is
$e^{ipX}(z)$. However, if $p$ is an integer or half-integer, then
the set of primary fields is bigger. At these values of the momentum
the primary fields form $SU(2)$ multiplets
$$\psi_{J, m}(z)\sim \left [H_-(z)\right ]^{J-m} :e^{iJX(z)}:\eqn\special$$
where $J=0, \half, 1, \ldots$, and $m=(-J, -J+1, \ldots, J)$.
We have introduced the
$SU(2)$ generators
$$\eqalign{& H_\pm (z)=\oint {du\over 2\pi i} :e^{\pm iX(u+z)}:\cr
& H_3 (z)=\oint {du\over 4\pi } \partial X(u+z)\ .\cr }
\eqn\eq$$
The connection with $SU(2)$ becomes apparent once we note that the
fields \special\ are the full set of primary fields in the compact
$c=1$ theory with the self-dual radius $R=2$, where the existence of the
$SU(2)$ current algebra is well known.
In the non-compact theory the operators \special\ form a subalgebra.
After performing the contour integrals in eq. \special,
$\psi_{J, m}$ become polynomials in derivatives of $X$.
Their dimension is $\Delta_J=J^2$. Upon gravitational dressing,
we obtain operators of dimension 1,
$$\eqalign{&\Psi_{J, m}^{(\pm)}(z)=\psi_{J, m} (z)
\exp (\epsilon_J^{(\pm)}\phi(z))\cr
&\epsilon_J^{(\pm)}=-1\pm J\ .\cr }
\eqn\eq$$
Here we have constructed the holomorphic part which is identical
in form to the open string vertex operators. For closed string
theory it must be multiplied by the antiholomorphic part.

The states \special\ are precisely those responsible for the
``leg poles'' at the discrete values of momenta found in
the tachyon Koba-Nielsen amplitudes [\AMP, \GKleb].
Our aim here is to determine
their self-interaction or, equivalently, the leading non-linear
piece of the $\beta$-function. It was realized in ``prehistoric times''
that the $\beta$-function is determined by the structure constants
in the operator product expansion [\sasha]. So, in order to determine
the quadratic piece of the $\beta$-function, one has to find
the structure constants in
$$\Psi_{J_1, m_1}^{(+)} (z)\Psi_{J_2, m_2}^{(+)} (0)=\ldots+
{1\over z} \sum_{J_3, m_3} F_{J_1, m_1, J_2, m_2}^{J_3, m_3}
\Psi_{J_3, m_3}^{(+)} (0)+\ldots
\eqn\fusion$$
and similarly for other signs of the Liouville dressing.
$F$ determines the open string cubic couplings. It is also clear
that $F$ is the structure constant in the algebra of charges
$ Q_{J, m}^{(+)}\sim \oint dz \Psi_{J, m}^{(+)}$.
First we need to determine which $J_3$ and $m_3$ contribute to the sum in
eq. \fusion.
It is clear that all operators on
the right-hand side depend on the zero modes through
$e^{i(m_1+m_2)X+(J_1+J_2-2)\phi}$. The only physical operator
with this property is, up to gauge transformations,
$\Psi_{J_1+J_2-1, m_1+m_2}^{(+)}$. Thus, the sum consists of a single
term with $J_3=J_1+J_2-1, m_3=m_1+m_2$.
Furthermore, the $SU(2)$ invariance requires
the structure constants $F$ to
be of the form
$$
F_{J_1, m_1, J_2, m_2}^{J_1+J_2-1, m_1+m_2}=
C_{J_1, m_1, J_2, m_2}^{J_1+J_2-1, m_1+m_2} g(J_1, J_2)
\eqn\Cleb$$
where $C$ are the Clebsch-Gordan coefficients and $g(J_1, J_2)$
is an unknown function. In other words, the $m$-dependence within
each $SU(2)$ multiplet is completely determined by the Clebsch-Gordan
coefficients. This can be shown directly by applying the $SU(2)$
generators $H_{\pm}$ to both sides of eq. \fusion.
For $J_3=J_1+J_2-1$, $m_3=m_1+m_2$, the coefficients assume the form
$$\eqalign{&C_{J_1, m_1, J_2, m_2}^{J_3, m_3}
={N(J_3, m_3)\over N(J_1, m_1) N(J_2, m_2)}
\times {J_2 m_1-J_1 m_2\over \sqrt {J_3 (J_3+1)}}\ ,\cr
& N(J, m)=\left [{(J+m)! (J-m)!\over (2J-1)!}\right ]^{1/2}\ .\cr
} \eqn\Clebschform$$
The final step is to determine the function $g(J_1, J_2)$.
For this purpose we consider the special values of $m_1$ and $m_2$
such that a direct calculation of the operator product is accessible.
Namely, take $m_2=J_2$ and $m_1=J_1-1$ so that
$$\Psi_{J_1, J_1-1}^{(+)} (z)\Psi_{J_2, J_2}^{(+)} (0)=\ldots+
z^{-1} F_{J_1, J_1-1, J_2, J_2}^{J_1+J_2-1, J_1+J_2-1}
\Psi_{J_1+J_2-1, J_1+J_2-1}^{(+)} (0)+\ldots
\eqn\spfusion$$
{}From the representation \special\ we get
$$\eqalign{&\sqrt{2J_1}\Psi_{J_1, J_1-1}^{(+)} (z)
\Psi_{J_2, J_2}^{(+)} (0)=\cr &
\oint {du\over 2\pi i}:e^{-iX(z+u)}:
\, :e^{iJ_1 X(z)}: \, :e^{iJ_2 X(0)}:
\, :e^{(J_1-1) \phi (z)}: \, :e^{(J_2-1) \phi(0)}:\cr
&=\oint {du\over 2\pi i} u^{-2J_1} (z+u)^{-2J_2}
z^{2(J_1+J_2-1)} \Psi_{J_1+J_2-1, J_1+J_2-1}^{(+)} (0)\cr
&=z^{-1}\oint {dx\over 2\pi i} x^{-2J_1} (1+x)^{-2J_2}
\Psi_{J_1+J_2-1, J_1+J_2-1}^{(+)} (0)\ .\cr }
\eqn\eq$$
Performing the integral, we find
$$ F_{J_1, J_1-1, J_2, J_2}^{J_1+J_2-1, J_1+J_2-1}
={(2J_1+2J_2-2)!\over \sqrt{2J_1} (2J_1-1)! (2J_2-1)!}
\ .\eqn\eq$$
Comparing this with eqs. \Cleb\ and \Clebschform, we obtain
$$ g(J_1, J_2)=-{\sqrt {J_1+J_2} (2J_1+2J_2-2)!\over
\sqrt {2J_1 J_2} (2J_1-1)! (2J_2-1)!}
\ . \eqn\eq$$
Thus, the structure constants assume a remarkably simple form
$$
F_{J_1, m_1, J_2, m_2}^{J_3, m_3}=\delta_{J_3, J_1+J_2-1}\delta_{m_3, m_1+m_2}
{\tilde N(J_3, m_3)\over \tilde N(J_1, m_1) \tilde N(J_2, m_2)}
(J_2 m_1-J_1 m_2)
\eqn\form$$
where $\tilde N(J, m)=-\sqrt{J/2}(2J-1)! N(J, m)$.
Now, changing the normalization of the special operators according to
$$\Psi_{J, m}^{(+)}\to {1\over \tilde N(J, m)}\Psi_{J, m}^{(+)}
\eqn\eq$$
we arrive at the following vertex operator algebra,
$$\Psi_{J_1, m_1}^{(+)} (z)\Psi_{J_2, m_2}^{(+)} (0)=z^{-1} (J_2 m_1-J_1 m_2)
\Psi_{J_1+J_2-1, m_1+m_2}^{(+)} (0) \ .
\eqn\finfusion$$
We can analyze the operator products involving $\Psi_{J, m}^{(-)}$
in complete analogy with the derivation above. We find
$$\eqalign{
&\Psi_{J_1, m_1}^{(-)}(z)\Psi_{J_2, m_2}^{(-)} (0)=z^{-1}\times 0\ ,\cr
&\Psi_{J_1, m_1}^{(+)}(z)\Psi_{J_2, m_2}^{(-)} (0)=z^{-1}\times 0\ ,
\qquad\qquad\qquad\qquad\qquad\qquad\qquad\qquad\qquad J_1\geq J_2+1\ ,\cr
&\Psi_{J_1, m_1}^{(+)} (z)
\Psi_{J_1+J_2-1, -m_1-m_2}^{(-)} (0)=z^{-1} (J_2 m_1-J_1 m_2)
\Psi_{J_2, -m_2}^{(-)} (0)\ ,
\qquad\qquad J_1<J_2+1
\ , \cr }
\eqn\mfusion$$
where we have redefined the normalization according to
$$\Psi_{J, m}^{(-)}\to \tilde N(J, m)\Psi_{J, m}^{(-)}
\ .\eqn\eq$$
If we define the charges
$$Q_{J, m}^{(\pm)}=  \oint {dz\over 2\pi i}\Psi_{J, m}^{(\pm)}(z)
\ ,\eqn\charges$$
then the vertex operator algebra \finfusion, \mfusion\ implies
that the non-zero commutators are given by
$$[Q_{J_1, m_1}^{(+)}, Q_{J_2, m_2}^{(+)}] =(J_2 m_1-J_1 m_2)
Q_{J_1+J_2-1, m_1+m_2}^{(+)} \eqn\walgebra$$
and
$$\eqalign{&
[Q_{J_1, m_1}^{(+)}, Q_{J_2, m_2}^{(-)}] =(J_2 m_1+J_1 m_2+m_1)
Q_{J_2-J_1+1, m_1+m_2}^{(-)}\ ,\cr
&J_2-J_1+1>0\ ,\qquad\qquad |m_1+m_2|\leq J_2-J_1+1\ .
\cr }\eqn\algebra$$
Eq. \walgebra\ is the algebra of area-preserving diffeomorphisms
of a plane [\Ed]. If we truncate to integer $J$, then we find the
``wedge''
sub-algebra of the well-known $w_{\infty}$ algebra [\bak].

This algebra gets naturally deformed as we turn on the cosmological
constant by adding the term $\lambda\int e^{-\phi}$ to the action.
Then there are additional terms on the right-hand side of
eq. \walgebra\ because insertions of the cosmological constant
violate the Liouville energy conservation. We get the structure
$$\eqalign{
&[Q_{J_1, m_1}^{(+)}, Q_{J_2, m_2}^{(+)}] =(J_2 m_1-J_1 m_2)
Q_{J_1+J_2-1, m_1+m_2}^{(+)}\cr
&+\sum_{n=1}^{J_1+J_2-1}\lambda^n a_n (J_1, m_1; J_2, m_2)
Q_{J_1+J_2-1-n, m_1+m_2}^{(+)}\cr
} \eqn\newalgebra$$
We have not calculated the extra structure constants
$a_n (J_1, m_1; J_2, m_2)$. Because of $SU(2)$ covariance
properties, we may conjecture that
the integer $J$ sub-algebra of
\newalgebra\ is isomorphic to ${\cal T}(\mu)$, the
enveloping algebra of $SU(2)$, for some value of $\mu$.
The ``wedge'' subalgebra of
$W_{\infty}$ is a special case of ${\cal T}(\mu)$ with $\mu=0$ [\PRS].
Another interesting case is ${\cal T}(\infty)$, which is the algebra
of area preserving diffeomorphisms
(or canonical transformations)
of a two-sphere $S^2$ [\PRS, \Berg]. If we describe points
on a sphere by three-vectors $\vec n$ of unit length, then
the transformation law is
$$\delta_W n_a=\epsilon_{abc} n_b {\partial W(\vec n)\over\partial n_c}
\ ,\eqn\diff$$
with $W(\vec n)$ being an arbitrary homogeneous polynomial
in $n_a$. It is not hard
to check that
$$\eqalign{&[\delta_{W_1}, \delta_{W_2}]=\delta_{W_3}\ ,\cr
&W_3= \{W_1, W_2\}=\epsilon_{abc} n_a {\partial W_1\over\partial n_b}
{\partial W_2\over\partial n_c}
\ .\cr }\eqn\sphdiff$$
This Poisson bracket induces $SU(2)$ algebra with the generators
$$ L_+=Y_{1,1}\sim x+iy\ ,\qquad
L_-=Y_{1,-1}\sim x-iy\ ,\qquad
L_0=Y_{1,0}\sim z\ .
\eqn\eq$$
The operators $Y_{l, m}$, which transform as higher spins
under this $SU(2)$, can be obtained by taking multiple Poisson brackets of
$Y_{1, -1}$ with the highest weight operators
$Y_{l, l}\sim\left(Y_{1, 1}\right)^l$.
The functions $Y_{l, m}$ found this way are the conventional
spherical harmonics. The algebra of area-preserving
diffeomorphisms of the sphere assumes the form
$$\eqalign{
&\{Y_{l_1, m_1}, Y_{l_2, m_2}\} ={M(l_1+l_2-1, m_1+m_2)
\over M(l_1, m_1) M(l_2, m_2)}(l_2 m_1-l_1 m_2)
Y_{l_1+l_2-1, m_1+m_2}\cr
& +\sum_{n=1} g_{2n}(l_1, l_2)
C_{l_1, m_1, l_2, m_2}^{l_1+l_2-1-2n, m_1+m_2}
Y_{l_1+l_2-1-2n, m_1+m_2}\ ,\cr
} \eqn\envalgebra$$
where $M(l, m)$ are normalization factors.
The construction sketched above can be used to show
that the algebra of area preserving diffeomorphisms of
$S^2$ is the classical ($\mu\to\infty$) limit of ${\cal T}(\mu)$ [\Berg].

The fusion rules \finfusion, \mfusion\ translate into the statement about
the leading term in the $\beta$-functions of the open string theory.
We introduce the Chan-Paton index $A$ in the adjoint
representation of some Lie group and associate coupling constants
$g_{J, m}^{(s), A}$ with vertex operators
$\Psi_{J, m}^{(s), A}$, where $s=\pm 1$ labels the
type of Liouville dressing. Since the vertex operators are on-shell,
the linear terms in the $\beta$-functions vanish. The quadratic terms are
$$\eqalign{
&\beta^{(+), A}_{J, m}=\half\sum_{J_1, m_1, J_2, m_2} (J_2 m_1-J_1 m_2)
f^{ABC} g^{(+), B}_{J_1, m_1} g^{(+), C}_{J_2, m_2}
\delta_{J, J_1+J_2-1}\delta_{m, m_1+m_2}+\CO(\lambda^2)\ , \cr
&\beta^{(-), C}_{J_2, -m_2}=\sum_{J_1, m_1} (J_2 m_1-J_1 m_2)
f^{ABC} g^{(-), A}_{J_1+J_2-1, -m_1-m_2} g^{(+), B}_{J_1, m_1}+
\CO(\lambda^2) \ , \cr
}\eqn\eq$$
where $f^{ABC}$ are the group structure constants.

The basic question of string theory concerns
the effective action for all the string modes.
Usually, the $\beta$-functions can be obtained as variations of an effective
action. In this case, it is not hard to see that
the $\beta$-functions for the discrete states can be expressed as
$$ \beta^{(+), A}_{J, m}={\partial S\over\partial g^{(-), A}_{J, -m} }\ ,
\qquad\qquad
\beta^{(-), A}_{J, m}={\partial S\over\partial g^{(+), A}_{J, -m} }\ ,
\eqn\eq $$
where the effective action is given by
\foot{If the $X$-coordinate is taken to
be non-compact, then we also need the conventional ``tachyon''
field $T(\phi, X)$. In this paper we will not consider interactions
of tachyons with the discrete states.}
$$ \eqalign {&S= S_2+S_3+S_4+\ldots \cr
&S_3=\half\sum_{J_1, m_1, J_2, m_2} (J_2 m_1-J_1 m_2)
f^{ABC} g^{(-), A}_{J_1+J_2-1, -m_1-m_2}
g^{(+), B}_{J_1, m_1} g^{(+), C}_{J_2, m_2} +\CO(\lambda^2)
\ . \cr }\eqn\cubic$$
One can check that $S_3$ correctly generates all the
three-point couplings of the discrete states. We have not yet
calculated the quartic and higher terms in the effective action.
In the next section, however, we will speculate on the correct
non-polynomial form.

To display the symmetries of the cubic term $S_3$,
it is convenient to assemble all the coupling constants into a field
\foot{Here we restrict
our discussion to states with integer
spin $J$. This is the full set of states in the compactified model
with radius $R=1$.}
$$\Phi(\phi, \vec n)=\sum_{s, A, J, m}T^A
g_{J, m}^{(s), A} M^s(J, m)Y_{J, m}(\vec n) e^{(s J-1)\phi}
\eqn\eq$$
where $T^A$ are the group generator matrices. As we see, this field
naturally depends on 3 coordinates.
Using eq. \envalgebra, we can write
the cubic action \cubic\ for $\lambda=0$ simply as
$$S_3\sim\int d\phi e^{2\phi} \int_{S^2} \Tr \left (
\epsilon_{abc} n_a \Phi {\partial \Phi\over\partial n_b}
{\partial \Phi\over\partial n_c}\right )
=\int d\phi e^{2\phi} \int_{S^2} d^2 x\Tr \left(
\epsilon_{ij} \Phi {\partial \Phi\over\partial x_i}
{\partial \Phi\over\partial x_j}\right)
\eqn\eff$$
where $x^i$ are coordinates on $S^2$, and
the $\Tr$ refers to the internal symmetry group.
Obviously, $S_3$ is invariant under the group
of area-preserving diffeomorphisms \diff.
For the case of non-compact string, we have to combine this
2+1 dimensional theory with a 1+1 dimensional theory of tachyons.
Thus, if Liouville coordinate is thought of as time, the
space is a direct sum of a two-sphere and a real line.
There is a precedent to this situation: in the sine-Gordon
model the base space is a direct sum of a real line and
a two-dimensional surface (a torus).

A natural extension
of \eff\ beyond the cubic order would be the WZNW term
$$S_{WZNW}=\int d\phi e^{2\phi} \int_{B^3} \Tr \left (\O^{-1} d\O\right )^3
\ ,\eqn\WZ$$
where $\O=e^\Phi$, and $B^3$ is the three-dimensional ball
whose boundary is $S^2$. We have not yet been able to
confirm this effective lagrangian by direct calculations.
There should also be a natural extension of the action \eff\ to
the case $\lambda\neq 0$.
Even if correct, the expression \WZ\ is written in a particular
``material'' gauge where all the vertex operators do not depend on the
derivatives of the Liouville field. Perhaps, the gauge-invariant
formulation will involve the Chern-Simons field $A(\phi, \vec n)$,
thus defining an open string field theory.

Now, we briefly discuss the case of closed string.
At the self-dual point we double $\Psi_{J, m}$ in a standard way,
defining the vertex operators
$$V_{J, m, m'}^{(s)}=
\int d^2 z\Psi_{J, m}^{(s)} (z)
\bar\Psi_{J, m'}^{(s)} (\bar z)\eqn\vertex$$
The couplings of these states are simply the products of the couplings
of $\Psi_{J, m}$ and
of $\bar\Psi_{J, m'}$. In other words, the $\beta$-functions
of the theory are given by
$$\eqalign{
&\beta^{(+)}_{J_1+J_2-1, m_1+m_2, m_1'+m_2'}=\half\sum
(J_2 m_1-J_1 m_2) (J_2 m_1'-J_1 m_2')
g^{(+)}_{J_1, m_1, m_1'} g^{(+)}_{J_2, m_2, m_2'}
+\CO(\lambda)\ , \cr } \eqn\eq$$
and similarly for $\beta^{(-)}$.
In order to write the effective action in a compact form, it is natural to
introduce not $S^2$, but $S^3$, so that the couplings are paired
with the functions ${\cal D}_{J, m, m'} (G)$ on the $SU(2)$ group manifold.
Thus, for closed strings we introduce a field
$$\Phi_c(\phi, G)=\sum_{J, m, m'}
{\cal D}_{J, m, m'}(G) \left (
e^{(J-1)\phi} g_{J, m, m'}^{(+)} n(J, m, m')
+e^{(-J-1)\phi} g_{J, m, m'}^{(-)} {J(J+1)\over n(J, m, m')}\right )
\eqn\eq$$
where $n(J, m, m')$ is the appropriate normalization factor.
Owing to the identity
$$\int_G {d\omega\over 8\pi^2} {\cal D}_{J_1, m_1, m_1'}
{\cal D}_{J_2, m_2, m_2'} {\cal D}_{J_3, m_3, m_3'}=
(J_1, m_1; J_2, m_2; J_3, m_3) (J_1, m_1'; J_2, m_2'; J_3, m_3')
\eqn\eq$$
where
$(J_1, m_1; J_2, m_2; J_3, m_3)$ is the $3j$-symbol, we can write the
cubic effective action for $\lambda=0$ as
$$S_3\sim \int d\phi e^{2\phi} \int_G d\omega\ \Phi_c^3
\ .\eqn\eq$$
This expression is invariant under volume preserving diffeomorphisms
of $S^3$.
Perhaps, in the general gauge this corresponds to some 3+1 dimensional
topological field theory. However, we are somewhat skeptical towards this
(and all other) traditional approaches to closed string field theory.
The reason is that it may happen that even the dimensionality
of the base space is background dependent and is a dynamical variable,
requiring a completely new approach.
We hope that the simple open string model considered above
will suggest the right direction.

Now, a few ``historical'' remarks are in order. The appearance of
the $w_\infty$ symmetry in the matrix models has been noted by
a number of authors [\avan].
In the continuous context it has recently been
found in ref. [\Ed]. We must say that, although we contemplated the
relations \fusion, \Cleb\ for a long time, we finished the
calculation only after E. Witten had informed us of his results.
We decided to publish this note because our methods are quite different
from his and, in our opinion, have their own merits.

\singlespace
\refout
\bye